\documentclass[cameraready]{Interspeech}

\title{The Hidden Cost of Pairwise Verification in Synthetic Speech Source Tracing}

\author[orcid=0000-0002-4717-1910, correspondingauthor]{Anton}{Firc}
\author[orcid=	0009-0001-1870-7704]{Zbyněk}{Lička}
\author[orcid=0009-0000-5722-0571]{Vojtěch}{Staněk}
\author[orcid=	0000-0002-9009-2193]{Kamil}{Malinka}

\address{
    Security@FIT, Brno University of Technology, Czech Republic
}

\email{\{ifirc, ilicka, istanek, malinka\}@fit.vut.cz}


\usepackage{comment}

\usepackage{amsmath}
\usepackage{enumitem}
\usepackage{subcaption}
\usepackage{graphicx}
\usepackage{textcomp}

\renewcommand{\paragraph}[1]{\vspace{0.5ex}\noindent\textbf{#1}\hspace{0.75em}}

\begin{document}

\maketitle

\begin{abstract}
Open-set source tracing is increasingly framed as a verification problem, motivating the use of pairwise metric-learning objectives from biometrics. We thus compare global anchoring and pairwise verification under matched backbones and a fixed data and epoch budget on MLAAD (in-domain) and STOPA (out-of-domain). In our runs, global anchoring yields lower in-domain error (8.61\% EER) than pairwise variants (12--15\% EER), even with rival mining and XLS-R finetuning. Because pairwise objectives optimize similarity directly, they concentrate variance into fewer embedding directions, reducing resolution among closely related generators. To test if this drives the drop, we impose a similar bottleneck to the globally supervised baseline, yet the baseline remains competitive. Together with an embedding-space analysis ($k_{99}$), these results suggest that the gap is not explained by dimensionality alone, but rather by the pairwise objective's shaping of the retained directions.
\end{abstract}

\section{Introduction}

As speech synthesis approaches evolved into a substantial security threat~\cite{biosigLidi,speech-biometric-practicalattack_firc22,deepfakesurvey_firc24,voiceAssitants}, audio deepfake detection provided a sufficient countermeasure against deepfake-related incidents.
Recently, Source Tracing has been explored to provide post-incident countermeasures by attributing the attack to the synthesizer used to create the deepfake.

Because the task involves tracing synthesizers unseen during training (i.e., an \emph{open-set} task), metric-learning objectives, particularly Siamese-style pairwise training, offer a compelling direction. These methods have proven useful for face and speaker recognition~\cite{Schroff_2015_CVPR,li2017deepspeakerendtoendneural}, operating on the assumption that optimizing local pairwise distances (i.e., \emph{local mining}) produces embeddings that generalize better to unseen classes. Recent benchmarks have begun to explore the feasibility of metric learning for source tracing under EER-centric evaluation~\cite{st-resnet-aasist-configurations-loss-sampling-embedding-size}.

We empirically test whether objective choices that work well in biometrics transfer to open-set source tracing. In this domain, discriminative cues can be subtle and synthesizer-specific, so objectives that emphasize stable margins may trade off fine-grained separability~\cite{stan25_interspeech}.
We therefore compare global anchoring (i.e., the baseline) and pairwise verification under matched training conditions on MLAAD~\cite{muller2024mlaad} and STOPA~\cite{firc25_interspeech} datasets.

In our runs, global anchoring~\cite{negroni25_interspeech} yields stronger in-domain verification than the tested pairwise variants, which is associated with a steeper embedding decay. 
As an ablation study, we include two controls, XLS-R finetuning and an explicit 10/13-dimensional bottleneck~\cite{st-resnet-aasist-configurations-loss-sampling-embedding-size} on the baseline.
Together, these results suggest that the gap cannot be explained by embedding dimensionality alone and point to objective-specific shaping of the retained embedding directions.

\noindent\textbf{Contributions.}
\begin{itemize}[leftmargin=*]
    \item We document that global anchoring outperforms verification-style pairwise training for open-set source tracing under matched backbones, protocols, and data/epoch budget.
    \item We show that this gap is not explained by backbone adaptation or dimensionality alone, but by objective-induced shaping of the embedding through controlled ablations and embedding-space analysis.
    \item We demonstrate a simple usage guideline for forensic source tracing: start with global anchoring, and use pairwise verification only if it demonstrably improves performance at low false-positive rates.
\end{itemize}

\section{From Classification to Verification}
\label{sec:related_work}

Early forensic works framed source tracing as a closed-set classification problem. Borrelli et al.~\cite{st-early-svm-randomforest-unknown-class_borrelli21} pioneered the use of SVMs to distinguish generator architectures, while subsequent studies decomposed the task into component-level analysis, classifying specific vocoders or acoustic models separately~\cite{st-early-components-no-unknown_zhu22, klein24_interspeech}. To improve robustness, recent approaches have integrated disentanglement mechanisms, such as removing speaker information via feature separation~\cite{st-adae-magnify-fingerprints-remove-speaker-info-fakesource-dataset} or feature fusion~\cite{st-trio-framework-fusion-strategy-canonical-correlation-analysis-paralinguistic-features_girish25,st-sota-comparison-finder-fusion-framework_phukan25,st-hydra-architecture-source-parsing-parameter-regression-benchmark_phukan25a}.

To overcome the rigidity of closed-set classification, recent work has shifted toward \emph{Source Verification}, which aims to determine whether two recordings share a common generator. This paradigm has emerged simultaneously across multiple studies: Negroni et al.~\cite{negroni25_interspeech} demonstrated a few-shot verification protocol, while parallel efforts have explored similar protocols across various architectures and training setups~\cite{firc25_interspeech,st-verification-siamese-mlp-scoring}. Several benchmarks have since explored verification-oriented objectives, including margin-based classification losses and prototypical metric learning~\cite{st-resnet-aasist-configurations-loss-sampling-embedding-size,st-verification-few-shot-identification-m2d-clap-internal-dataset}. Most notably, Koutsianos et al.~\cite{st-resnet-aasist-configurations-loss-sampling-embedding-size} provide a systematic comparison on MLAAD, demonstrating that embeddings trained with global angular margins~\cite{asv-softmax-losses-comparison_xiang19} can be compressed to as few as 10 dimensions without degrading performance. 
This challenges the assumption that forensic traces strictly require high-dimensional representations and suggests that the \emph{optimization path}
is the determining factor for representation quality.

\paragraph{Positioning.}
Our work examines how local mining (i.e., pairwise training) performs for synthetic speech attribution compared to global anchoring (i.e., cross-entropy).
In our experiments, pairwise training is associated with a much steeper embedding decay (Section~\ref{sec:manifold_analysis}) and weaker in-domain verification performance than global anchoring.
We also include an ablation study in which we enforce a low-dimensional embedding (10/13 dims) under global anchoring, which remains competitive on MLAAD.
This contrast suggests that the outcome is not explained by dimensionality alone, and it motivates studying how different objectives select and suppress directions in the representation space.

\section{Experimental Framework}
\label{sec:framework}

\subsection{Hypothesis and Comparison Strategy}

We study how training objectives affect verification performance and the geometry of representation for open-set source tracing. Motivated by the success of pairwise learning in biometrics~\cite{Schroff_2015_CVPR,li2017deepspeakerendtoendneural}, we test whether pairwise objectives improve generalization or trade off fine-grained resolution among closely related generators. We contrast two training strategies:

\begin{itemize}[leftmargin=*]
    \item \textbf{Global Anchoring (Baseline)}~\cite{negroni25_interspeech}: Using closed-set classification as a proxy for global class-based learning~\cite{asv-softmax-losses-comparison_xiang19}, we force samples to align with learned class centers. This approach solves a multi-class separation problem and can encourage compact within-generator embeddings.
    \item \textbf{Local Mining (Target):} Using pairwise Siamese constraints to optimize relative distances. This approach directly optimizes similarity between target and non-target pairs and can emphasize large margins between classes. In our analysis, we examine whether this training setup is associated with a more concentrated embedding space and reduced resolution for closely related generators.
\end{itemize}

This comparison isolates global multi-class from local pairwise constraints under matched architectures and a fixed data and epoch budget.

\subsection{Data Protocols}
\label{sec:task_data_protocols}

We use MLAADv8~\cite{muller2024mlaad} for in-domain development and evaluation, and STOPA~\cite{firc25_interspeech} for out-of-domain (OOD) evaluation.

\noindent\textbf{Pairwise development.}
Pairwise source verification does not come with a single standard protocol in MLAAD, so we define a development trial list for tuning. Our dev list contains 97k trials (20k target, 77k non-target) and is kept fixed across runs\footnotemark.

\footnotetext{\scriptsize\url{https://github.com/Security-FIT/hidden-cost-pairwise-verification}}

\noindent\textbf{Claim-based evaluation.}
Following the claim-based setup of Negroni et al.~\cite{negroni25_interspeech}, each Generator ID defines a claim $\mathcal{C}$.
For each claim, we enroll $R$ utterances sampled uniformly from the evaluation set and score all remaining utterances against all claims.
We report $R=1$ on both MLAAD and STOPA for cross-dataset consistency, and we additionally report $R=5$ on MLAAD to compare with prior work.
For $R>1$, we select the highest similarity from enrolled utterances.
On MLAAD with XLS-R finetuning, our \textit{baseline} system attains $7.99\%$ EER at $R=1$ and $5.50\%$ EER at $R=5$ (mean over 3 seeds). This is comparable to the original report (Table 1 in~\cite{negroni25_interspeech}), which claims $R=5$ EERs of $8.3\%$ (AASIST) and $4.8\%$ (ResNet).

\subsection{System Architecture}
To attribute performance differences primarily to the training objective, we keep the backbone and pooling backend fixed within the main comparisons. All systems share the same XLS-R backbone (Wav2Vec 2.0 XLS-R~\cite{Babu2021}, 300M), unless stated otherwise, the backbone is frozen. We evaluate two pooling strategies: the graph-based backend AASIST~\cite{aasist} and Multi-Head Factorized Attention (MHFA)~\cite{mhfa}.
We release the full training and evaluation code to facilitate future research\footnotemark[\value{footnote}].

\subsection{Training Objectives}
\label{sec:training_objectives}

\noindent\textbf{1. Global anchoring (Baseline).}
We re-implement the attribution-based verification framework established by Negroni et al.~\cite{negroni25_interspeech}. This approach treats open-set verification as a representation learning problem via closed-set classification. The model projects the pooled embedding $h$ to class logits via a linear layer and is optimized using Softmax Cross-Entropy over the $N=24$ training generators. At inference, we extract embeddings from the penultimate layer and compute verification scores using Cosine Similarity. This strategy implicitly structures the dimensions around global class centers.

\noindent\textbf{2. Pairwise Verification.}
Pairwise systems replace the classification head with a fusion module that maps an embedding pair $(h_a,h_b)$ to a scalar similarity score, estimating the probability that the two samples were generated by the same generator. We compare four trial selection regimes:

\begin{itemize}[leftmargin=*]
    \item \textbf{Intermediate (Random):} A baseline regime sampling anchor-positive pairs against random negatives (1:1 ratio), ensuring broad dimension coverage but lacking boundary focus.
    \item \textbf{Hard-Negative Mining (Latent):} We select the numerically hardest non-targets for each anchor using a teacher model.
    \item \textbf{Directional (Coverage-Driven):} A geometric strategy that selects anchors via k-means clustering to maximize coverage, forming local neighborhoods of trials within controlled similarity bands to combine global coverage with local structure~\cite{bennouna2025dataenablesoptimaldecisions}.
    \item \textbf{Rival Mining (Metadata-Guided):} A strategy that explicitly targets difficult pairs using metadata. We mine \textit{Structural Rivals} (overlapping generator architectures, e.g., \texttt{Bark} vs. \texttt{Bark-small}) to force resolution of quantization artifacts, and \textit{Disentanglement Rivals} (same-speaker/different-generator pairs) to penalize reliance on speaker identity. In each batch, we replace 50\% of random non-targets with rival pairs, keeping the 1:1 target:non-target ratio fixed.
\end{itemize}

\paragraph{Training budgets and controls.}
We hold the data, epochs, and core pipeline (backbone, pooling head, optimizer family) fixed across objectives. Because pairwise systems train on sampled pairs, per-objective computation is not strictly equal.

\paragraph{Architecture selection.}
On the MLAAD development split, we swept two pooling backends and six pairwise scoring heads (3 seeds).
MHFA was the most stable and achieved the best average in-domain verification, so we use \texttt{XLS-R+MHFA} as the default.
For pairwise training, we adopt the \texttt{FFCosine} head, a learned affine transform of cosine similarity: $s=w\cos(h_a,h_b)+b$.
Full grid results are in the supplementary material.

\section{Experiments and Results}

\begin{table*}[t]
    \centering
    \caption{\textbf{Global vs.\ pairwise objectives with bottleneck and backbone controls.}
    Mean metrics over $N=3$ seeds with $R{=}1$ on MLAAD (in-domain)\textsuperscript{$\dagger$} and STOPA (OOD)\textsuperscript{$\ddagger$}.
    Best values are highlighted within each system family (Global vs.\ Pairwise).}
    \label{tab:main_results}
    \setlength{\tabcolsep}{2.2pt}

    \resizebox{0.95\linewidth}{!}{%
    \begin{tabular}{l c c c c c c c}
        \toprule
        & \multicolumn{4}{c}{\textbf{MLAAD (In-Domain)}} & \multicolumn{3}{c}{\textbf{STOPA (OOD)}} \\
        \cmidrule(lr){2-5} \cmidrule(lr){6-8}
        \textbf{System} &
        \textbf{EER (\%$\downarrow$)} &
        \textbf{nDCF$_{0.01}$ ($\downarrow$)} &
        \textbf{TPR@0.01\% (\%$\uparrow$)} &
        \textbf{TPR@0.1\% (\%$\uparrow$)} &
        \textbf{EER (\%$\downarrow$)} &
        \textbf{TPR@0.01\% (\%$\uparrow$)} &
        \textbf{TPR@0.1\% (\%$\uparrow$)} \\
        \midrule
        \textbf{Global (CE)} &
        8.61 $\pm$ 0.29 &
        0.90 $\pm$ 0.06 &
        4.42 $\pm$ 3.51 &
        19.29 $\pm$ 5.53 &
        30.81 $\pm$ 3.19 &
        \textbf{0.16} $\pm$ 0.10 &
        1.18 $\pm$ 0.54 \\
        \quad + XLS-R finetune &
        7.99 $\pm$ 0.45 &
        0.87 $\pm$ 0.06 &
        5.50 $\pm$ 4.47 &
        21.83 $\pm$ 5.76 &
        30.77 $\pm$ 2.78 &
        \textbf{0.16} $\pm$ 0.09 &
        \textbf{1.25} $\pm$ 0.53 \\
        \quad + emb bottleneck (10) &
        \textbf{7.05} $\pm$ 0.79 &
        \textbf{0.83} $\pm$ 0.01 &
        6.82 $\pm$ 3.05 &
        \textbf{26.79} $\pm$ 1.20 &
        31.63 $\pm$ 0.85 &
        0.06 $\pm$ 0.01 &
        0.54 $\pm$ 0.11 \\
        \quad + emb bottleneck (13) &
        8.84 $\pm$ 1.03 &
        \textbf{0.83} $\pm$ 0.03 &
        \textbf{7.56} $\pm$ 3.54 &
        26.02 $\pm$ 2.82 &
        \textbf{27.74} $\pm$ 5.14 &
        0.07 $\pm$ 0.01 &
        0.59 $\pm$ 0.14 \\
        \midrule
        \multicolumn{8}{l}{\textit{Pairwise (BCE)}} \\
        Intermediate (scratch) &
        14.92 $\pm$ 2.43 &
        0.99 $\pm$ 0.01 &
        1.57 $\pm$ 0.35 &
        8.97 $\pm$ 2.49 &
        29.54 $\pm$ 1.47 &
        0.13 $\pm$ 0.05 &
        0.93 $\pm$ 0.15 \\
        Hard-mined &
        15.06 $\pm$ 1.97 &
        0.99 $\pm$ 0.01 &
        1.58 $\pm$ 0.28 &
        8.95 $\pm$ 2.12 &
        30.32 $\pm$ 2.18 &
        0.12 $\pm$ 0.05 &
        0.89 $\pm$ 0.18 \\
        Directional &
        15.12 $\pm$ 1.97 &
        0.99 $\pm$ 0.01 &
        1.51 $\pm$ 0.22 &
        8.65 $\pm$ 2.21 &
        30.31 $\pm$ 2.14 &
        0.12 $\pm$ 0.04 &
        0.87 $\pm$ 0.16 \\
        Rival mining &
        14.22 $\pm$ 1.99 &
        \textbf{0.98} $\pm$ 0.01 &
        \textbf{2.33} $\pm$ 0.64 &
        \textbf{10.65} $\pm$ 2.89 &
        29.38 $\pm$ 1.44 &
        0.14 $\pm$ 0.05 &
        0.99 $\pm$ 0.17 \\
        Rival + XLS-R finetune &
        \textbf{12.39} $\pm$ 2.82 &
        0.99 $\pm$ 0.01 &
        1.41 $\pm$ 0.61 &
        10.31 $\pm$ 1.27 &
        \textbf{28.48} $\pm$ 1.17 &
        \textbf{0.16} $\pm$ 0.06 &
        \textbf{1.06} $\pm$ 0.17 \\
        \bottomrule
    \end{tabular}%
    }

    \vspace{2pt}
    \par\noindent
    {\scriptsize\raggedright
    $\dagger$~nDCF$_{0.01}$ is normalized so reject-all gives 1 under $P_{\mathrm{tar}}{=}0.01$, $C_{\mathrm{miss}}{=}1$, $C_{\mathrm{fa}}{=}1$. Reported with fixed-FPR TPR to resolve strict low-FPR performance. \par
    $\ddagger$~STOPA: DCF is omitted because it saturates near reject-all under the same prior. We report EER and fixed-FPR TPR instead. \par
    }
\end{table*}

\subsection{Establishing the Pairwise Baseline}
\label{ssec:baseline_selection}

Unless stated otherwise, we use the pairwise defaults tuned on MLAAD-dev (3 seeds), namely \textit{intermediate} sampling and \texttt{XLS-R+MHFA}+\texttt{FFCosine}, and keep the training budget fixed thereafter (full sweeps in the supplement).

\subsection{Benchmark: Global vs. Pairwise Optimization}
\label{sec:benchmark_results}

Across the configurations we tested, global anchoring yields the lowest MLAAD error and the strongest recall at strict operating points (Table~\ref{tab:main_results}, left). Pairwise systems benefit from rival mining and from XLS-R finetuning, but they remain worse than the global baselines on MLAAD in our runs.

\paragraph{Generalization and Domain Shift.}
All systems degrade substantially under domain shift on STOPA (Table~\ref{tab:main_results}, right). At 0.1\% FPR, TPR remains below roughly 1--1.3\% across methods in our runs, so differences between objectives are small in the strict forensic regime. For context, our best STOPA result reaches $27.74\%$ EER (Table~\ref{tab:main_results}), improving over the pilot baseline reported with STOPA (AASIST CM at $39.15\%$ pooled EER on known-attack trials - Table~4 in~\cite{firc25_interspeech}), although strict low-FPR TPR remains low for all methods. Because DCF saturates near the reject-all baseline on STOPA under the same rare-target profile, we focus on EER and fixed-FPR TPR and treat the STOPA OOD ordering as suggestive rather than conclusive.

\begin{figure}
    \centering
    \includegraphics[width=0.9\linewidth]{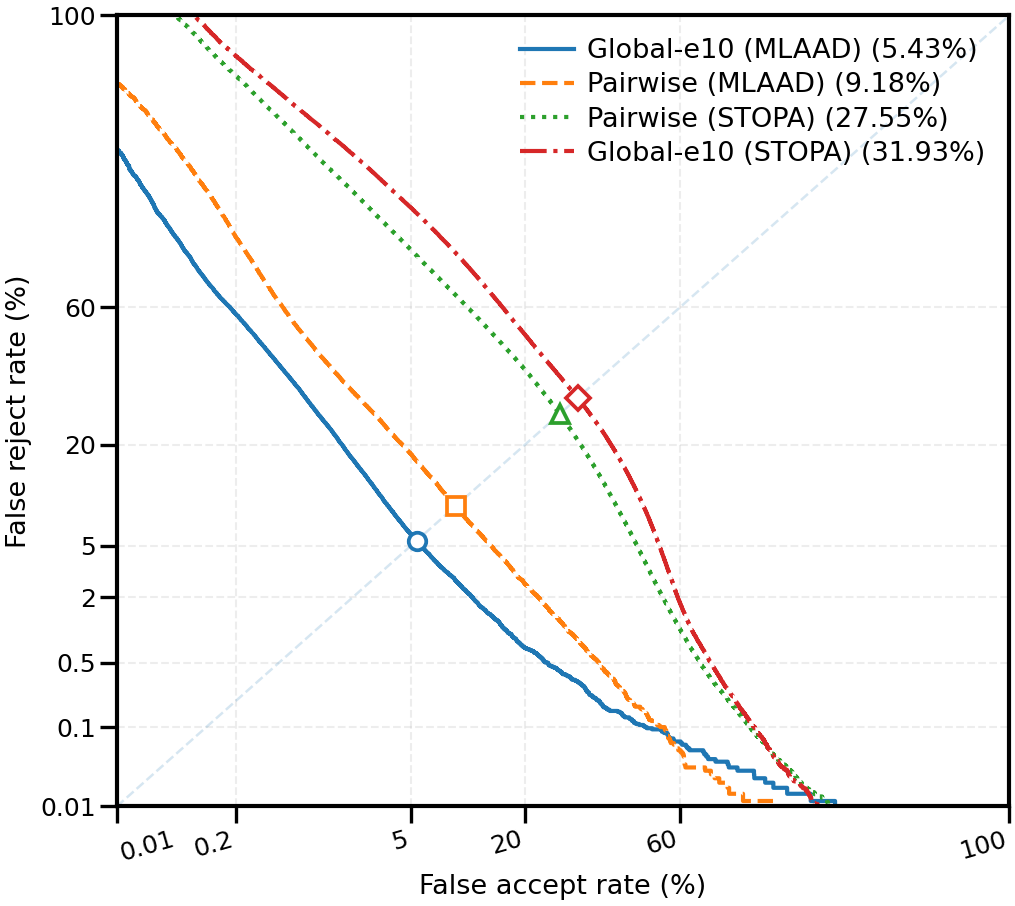}
    \caption{DET curves for the best Global and Pairwise checkpoints on MLAAD (left) and STOPA (right).
    Differences at strict operating points are dominated by low-FPR tail overlap.}
    \label{fig:det_global_pairwise}
\end{figure}

Figure~\ref{fig:det_global_pairwise} plots DET curves for the best Global and Pairwise systems on MLAAD and STOPA, highlighting behavior at strict low-FPR operating points.

\paragraph{Additional pairwise ablations.}
Beyond the reported BCE pairwise variants, we tested additional objectives and initialization variants, including raw cosine similarity with a margin loss and initializing pairwise training from global (CE) baselines instead of training from scratch. These margin-based cosine objectives performed worse than the BCE pairwise variants on MLAAD and did not improve STOPA, and \textit{none of these ablations changed the overall ordering}; we report them in the supplementary material.

\subsection{Embedding Topology Analysis}
\label{sec:manifold_analysis}

To better characterize the difference between objectives, we analyze the embedding decay of the embeddings using $k_{99}$, the number of principal components required to explain 99\% of the variance. As a variance-based measure, $k_{99}$ quantifies the concentration of the embedding distribution, but it does not directly indicate how much task-relevant information is preserved.

\begin{figure}
    \centering
    \includegraphics[width=\linewidth]{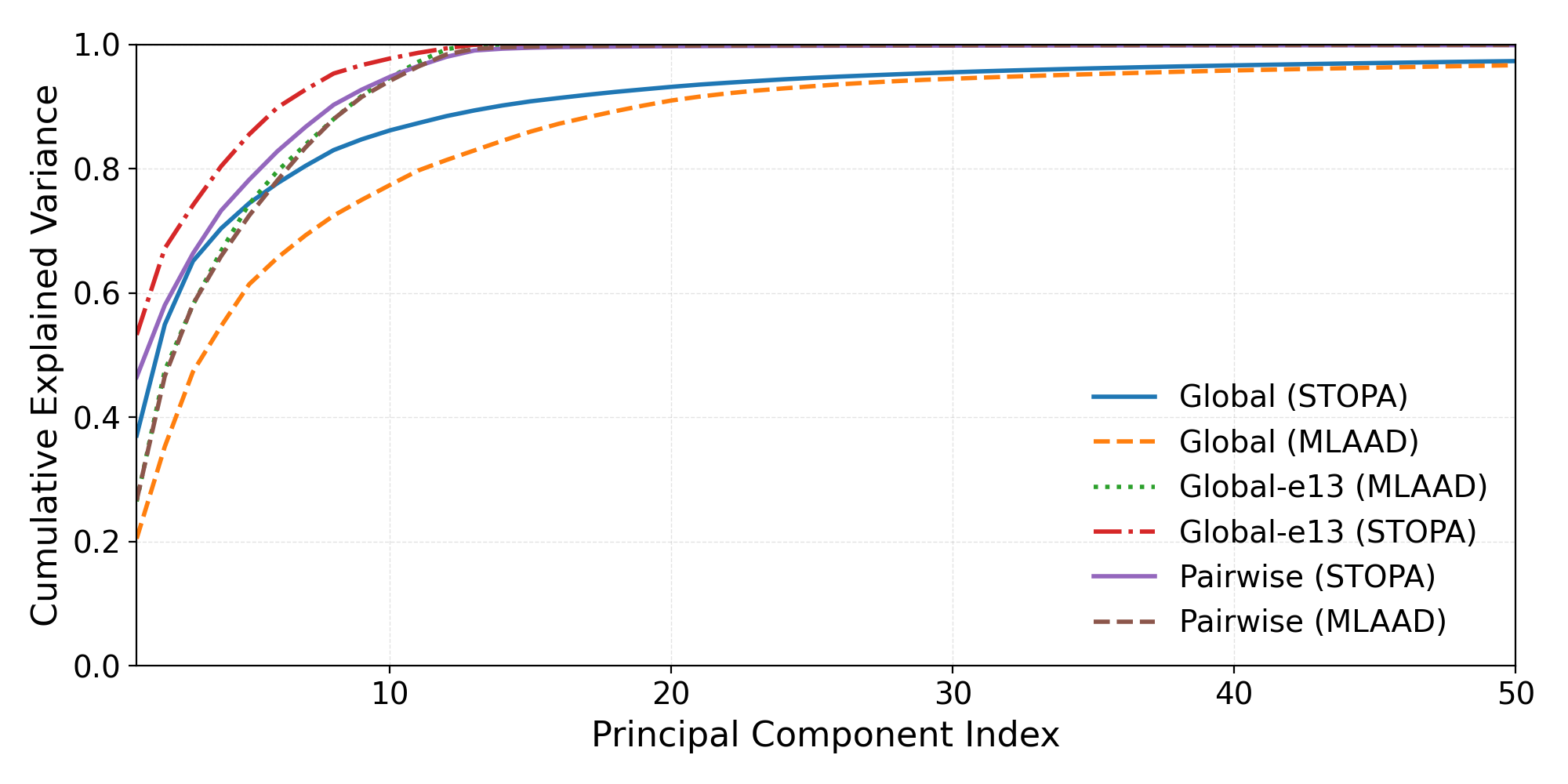}
    \caption{Cumulative variance analysis demonstrating dimensionality collapse.}
    \label{fig:cumvar}
\end{figure}

Figure~\ref{fig:cumvar} shows that the optimization paradigm is associated with a markedly different embedding space. The global anchoring objective exhibits a slower decay and a higher $k_{99}$ (e.g., $k_{99}\approx121$ in our MLAAD runs), whereas the pairwise objective is associated with a much steeper decay (e.g., $k_{99}\approx13$), even when the backbone is finetuned.

Importantly, low dimensionality alone does not appear sufficient to explain the performance drop. Prior work reports effective compression under global supervision~\cite{st-resnet-aasist-configurations-loss-sampling-embedding-size}, and in our bottleneck runs, a globally supervised 10--13 dimensional embedding remains competitive on MLAAD (Table~\ref{tab:main_results}). This contrast suggests that the relevant factor is not only \emph{how many} directions are retained, but also \emph{which} directions are emphasized.

We therefore treat embedding decay as a diagnostic rather than a complete explanation. Figure~\ref{fig:score_dist} shows that, in general, pairwise systems push target and non-target means further apart, but also yield substantially wider score distributions with heavier tails. This increased spread widens tail overlap at strict thresholds, which is exactly the regime that matters for forensic operations, and it is reflected in the degraded TPR at a fixed low FPR (e.g., TPR@0.1\%) in our setting.

\subsection{Fine-Grained Error Analysis}
\label{sec:failure_modes}

To better understand the source of the performance disparity at strict operating points (FPR $< 1\%$), we examined the specific pairs contributing to false acceptance errors. The breakdown of ``impostor pairs'' (Table~\ref{tab:tail_errors_expanded}) reveals distinct behaviors depending on the acoustic similarity of the sources.

\noindent\textbf{1. Shared Limitations (Digital Twins).}
For pairs sharing identical architectures and training data (e.g., \texttt{VITS} vs. \texttt{VITS-Neon}), both objectives exhibit high error rates. As indicated by the binary probe experiments (Section~\ref{sec:digital_twins}), these sources appear topologically overlapping in the XLS-R feature space. The inability to distinguish them reflects a limitation of the extractor rather than a limitation of the training objective.

\noindent\textbf{2. Loss of Resolution (Architectural Cousins).}
A critical divergence appears for systems that share an architecture but differ in configuration, such as \texttt{Multi-Dataset-Bark} vs. \texttt{Bark-Small}. While the Global Model retains sufficient discriminability to separate these variants in our setting (480 errors), the Pairwise Model exhibits a nearly threefold increase in confusion (1,269 errors). This pattern is consistent with the steeper embedding decay observed under pairwise training (Section~\ref{sec:manifold_analysis}), which may reduce resolution for subtle cues.

\begin{figure}
    \centering
    \includegraphics[width=\linewidth]{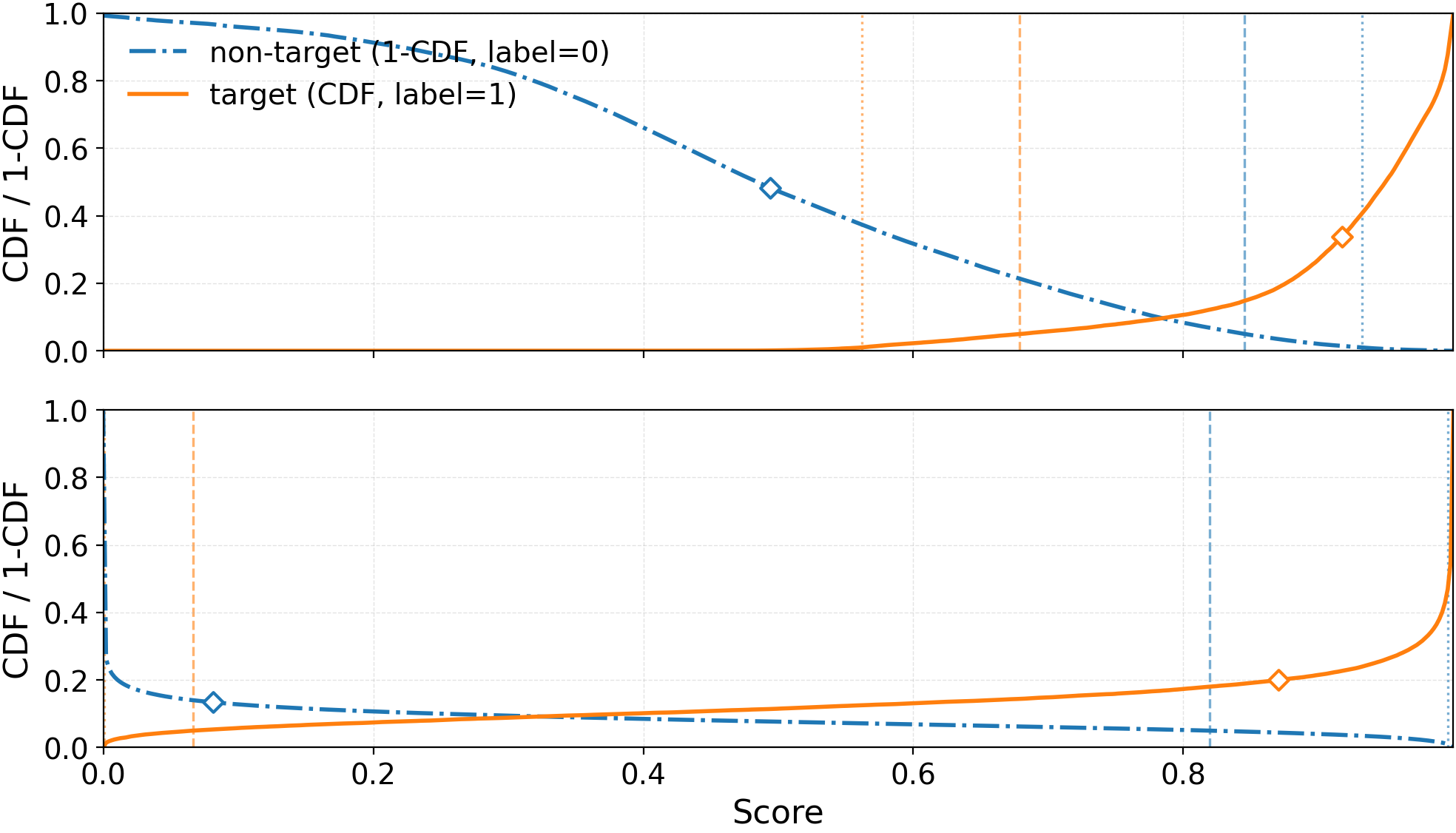}
    \caption{Score CDFs on MLAAD for \textbf{Global (CE) embneck13} (top) and \textbf{Pairwise Rival + XLS-R finetune} (bottom). We plot target $\mathrm{CDF}$ and non-target $1-\mathrm{CDF}$. Diamonds mark means; vertical lines mark 95\% and 99\% quantiles.}
    \label{fig:score_dist}
\end{figure}

We do not claim that $k_{99}$ alone explains these errors. However, the more concentrated embedding space under pairwise training suggests that fewer directions account for most embedding variance, which could make it harder to encode subtle configuration cues. This interpretation is also consistent with the score overlap between positive and negative classes shown in Figure~\ref{fig:score_dist}.

\begin{table}
    \centering
    \caption{False Acceptance Breakdown (FPR=0.1\%). Top false-accept generator pairs for the Global baseline and two Pairwise variants.}
    \label{tab:tail_errors_expanded}
    \setlength{\tabcolsep}{3pt}
    \resizebox{\columnwidth}{!}{%
    \begin{tabular}{l l c c c}
        \toprule
        & & \multicolumn{3}{c}{\textbf{False Accept Count $\downarrow$}} \\
        \cmidrule(lr){3-5}
        \textbf{Error Type} & \textbf{Generator Pair} & \textbf{Baseline} & \textbf{Hardmined} & \textbf{Rival} \\
        \midrule
        \textbf{Model Size} & \texttt{Bark} vs \texttt{Multi-Bark} & 480 & 1,269 & 881 \\
        \textbf{Model Size} & \texttt{Bark-Small} vs \texttt{Multi-Bark} & 549 & \textbf{113} & \textbf{75} \\
        \textbf{Architecture} & \texttt{Tacotron2} vs \texttt{XTTS\_v2} & 366 & $<10^\dag$ & $<10^\dag$ \\
        \textbf{Language} & \texttt{MMS-Hun} vs \texttt{MMS-Swe} & $<20^\dag$ & 84 & 122 \\
        \textbf{Language} & \texttt{VITS-Lt} vs \texttt{VITS-Mt} & $<13^\dag$ & 47 & 73 \\
        \bottomrule
    \end{tabular}
    }
    \vspace{2pt}
    \footnotesize{\emph{$^\dag$Indicates the pair did not appear in the top-k error list for that model.}}
\end{table}

\subsection{Separability of Generator Variants}
\label{sec:digital_twins}

We utilized binary classifiers to test the separability of specific generator pairs (Table~\ref{tab:probe_results}).
Consistent with recent reports on the difficulty of same-architecture attribution~\cite{st-resnet-aasist-configurations-loss-sampling-embedding-size,st-verification-few-shot-identification-m2d-clap-internal-dataset}, we find that pairs sharing both architecture and training data (e.g., \texttt{VITS}/\texttt{VITS-Neon}, \texttt{Parler-L}/\texttt{Parler-S}) are effectively indistinguishable (EER $\approx 50\%$). This confirms that standard XLS-R features are invariant to these specific capacity changes.

Crucially, our probe reveals that this limitation is driven by corpus and speaker overlap. While separating \texttt{Suno Bark} from its smaller variant (same data) proves difficult ($39.0\%$ EER), distinguishing \texttt{Bark-Small} from \texttt{Multi-Dataset Bark} is trivial ($2.0\%$ EER). 
This contrast mirrors the findings of Stan et al.~\cite{stan25_interspeech}, who demonstrated that source tracing performance drops significantly (from F1 $\approx 0.99$ to $0.74$) when speaker identity is held constant.
Together, these results imply that the discriminative power of current backbones is reliant on corpus-specific factors, such as channel characteristics and speaker identity, which often overshadow the subtle traces of the generative architecture itself.

\begin{table}
    \centering
    \caption{Binary probe results: EER (\%) for classifiers trained on specific generator pairs.}
    \label{tab:probe_results}
    \setlength{\tabcolsep}{3pt}
    \resizebox{\columnwidth}{!}{%
    \begin{tabular}{l l c}
        \toprule
        \textbf{Condition} & \textbf{Model Pair} & \textbf{EER (\% $\downarrow$)} \\
        \midrule
        Cross-Lingual & \texttt{MMS-Deu} vs \texttt{MMS-Hun} & $<1.0$ \\
        Cross-Corpus & \texttt{Multi-Dataset-Bark} vs \texttt{Suno Bark-Small} & $2.0$ \\
        \midrule
        Digital Twin & \texttt{Suno Bark} vs \texttt{Suno Bark-Small} & $39.0$ \\
        Digital Twin & \texttt{Parler-Large} vs \texttt{Parler-Mini} & $49.0$ \\
        Digital Twin & \texttt{VITS-LJ} vs \texttt{VITS-Neon-LJ} & $50.0$ \\
        \bottomrule
    \end{tabular}
    }
\end{table}

\section{Conclusion}
\label{sec:conclusions}

Across our experiments, global supervision remains a strong baseline for open-set synthetic speech attribution: it achieves the best in-domain verification on MLAAD, while the tested pairwise variants improve with mining and SSL finetuning but do not match it. Diagnostics of the learned representations suggest that the gap is not explained by embedding dimensionality alone: pairwise training concentrates variance into fewer directions and yields heavier-tailed score distributions, degrading separation at strict forensic thresholds, whereas low-dimensional bottlenecks trained under global supervision remain competitive. Overall, our results highlight a key objective trade-off between mean separation and tail behavior, and they motivate objectives that improve low-FPR separability without losing fine-grained generator cues.

\paragraph{Limitations.}
Our conclusions reflect the pairwise setting evaluated here and the XLS-R backbone with the tested pooling heads.
Stronger or differently tuned metric-learning objectives (e.g., supervised contrastive or proxy-based losses) may yield different trade-offs.
On STOPA, all methods degrade sharply, so we view the OOD ordering as indicative rather than definitive.

\section{Acknowledgements}
This work was partially supported by the Brno University of Technology (internal project FIT-S-23-8151) and the Ministry of Education, Youth and Sports of the Czech Republic through the e-INFRA CZ (ID:90254)

\section{Generative AI Use Disclosure}

During the preparation of this work, the authors used Generative AI Models (specifically Google Gemini, ChatGPT, and Grammarly) for language editing and text refinement. The authors reviewed and edited the output as needed and take full responsibility for the publication's content.

\bibliographystyle{IEEEtran}
\bibliography{mybib}

\end{document}